\documentclass[journal]{IEEEtran}
\usepackage{color}
\usepackage{graphicx}
\usepackage{siunitx}
\usepackage[utf8]{inputenc}

\usepackage{cite}

\ifCLASSINFOpdf
\else
\fi
\usepackage{amsmath}
\usepackage{url}

\newcommand{\openai}{OpenAI Gym}
\newcommand{\RL}{RL}
\newcommand{\gymem}{GEM}
\newcommand{\ML}{ML}
\newcommand{\PMSM}{PMSM}
\newcommand{\DDPG}{DDPG}
\newcommand{\DQN}{DQN}
\newcommand{\DNN}{DNN}
\newcommand{\action}{\bm{a}}
\newcommand{\observ}{\bm{o}}
\newcommand{\state}{\bm{s}}
\usepackage{booktabs}
\usepackage{bm}
\usepackage{subcaption}

\newcommand{\de}{\mathrm{d}}
\newcommand{\noise}{\sigma}
\newcommand{\MAE}{MAE}

\begin{document}
%
\title{Towards a Reinforcement Learning Environment Toolbox for Intelligent Electric Motor Control}
%
%
%

\author{Arne~Traue,
        Gerrit~Book,
        Wilhelm~Kirchgässner,~\IEEEmembership{Member,~IEEE}
        and Oliver~Wallscheid,~\IEEEmembership{Member,~IEEE}
\thanks{A. Traue, G. Book, W. Kirchgässner and O. Wallscheid are with the Department of Power Electronics and Electrical Drives at Paderborn University, Germany.
	e-mail: \{trauea, gbook\}@mail.uni-paderborn.de, \{kirchgaessner, wallscheid\}@lea.uni-paderborn.de}}
\maketitle

\begin{abstract}
Electric motors are used in many applications and their efficiency is strongly dependent on their control.
Among others, PI approaches or model predictive control methods are well-known in the scientific literature and industrial practice.
A novel approach is to use reinforcement learning (\RL) to have an agent learn electric drive control from scratch merely by interacting with a suitable control environment.
\RL~achieved remarkable results with super-human performance in many games (e.g. Atari classics or Go)  and also becomes more popular in control tasks like cartpole or swinging pendulum benchmarks. In this work, the open-source Python package \mbox{gym-electric-motor (\gymem)} is developed for ease of training of \RL-agents for electric motor control. Furthermore, this package can be used to compare the trained agents with other state-of-the-art control approaches. It is based on the \openai~framework that provides a widely used interface for the evaluation of \RL-agents. The initial package version covers different DC motor variants and the prevalent permanent magnet synchronous motor as well as different power electronic converters and a mechanical load model. Due to the modular setup of the proposed toolbox, additional motor, load, and power electronic devices can be easily extended in the future. Furthermore, different secondary effects like controller interlocking time or noise are considered. An intelligent controller example based on the deep deterministic policy gradient algorithm which controls a series DC motor is presented and compared to a cascaded PI-controller as a baseline for future research. Fellow researchers are encouraged to use the framework in their \RL~investigations or to contribute to the functional scope (e.g. further motor types) of the package.

\end{abstract}


%
\IEEEpeerreviewmaketitle

\section{Introduction}
%
%
%
%
\IEEEPARstart{E}{lectric} motor control has been an important topic in research and industry for decades, and a lot of different strategies have been invented, e.g. PI-controller and model predictive control (MPC) \cite{Linder.2010}. The latter methods require an accurate model of the system. Based on this, the next control action is calculated through an online optimization over the next time steps \cite{Gorges.2017}. Typical challenges when implementing MPC algorithms in drive systems are the computational burden due to the real-time optimization requirement and plant model deviations leading to inferior control performance during transients and in steady-state.

Furthermore, many breakthroughs in the recent years have been possible due to machine learning (\ML) and especially deep neural networks (\DNN). An example is the field of computer vision. After AlexNet \cite{Krizhevsky.2017} has won the ImageNet classification challenge in 2012, \DNN~have dominated research in many high level image processing tasks. Even reinforcement learning (\RL) was influenced by \DNN. New algorithms like deep-Q-learning (\DQN) \cite{Mnih.2015} and deep deterministic policy gradient (\DDPG) \cite{Lillicrap.992015} have been established. A famous example is the \RL-agent AlphaGo \cite{Silver.2017} which has beaten the currently best human player in the game of Go recently, and sparked new interest in the field of self-learned decision-making.
In the past years, \RL~has been applied to many control tasks like the inverse pendulum \cite{Panyakaew.2018}, the double pendulum \cite{Hesse.2018} or the cartpole problem \cite{Fremaux.2013}, and the application in electric power systems is also investigated \cite{Glavic.2017}.

Applying \RL~to electric motor control is an emerging approach \cite{Schenke.}. In contrast to MPC, \RL~control methods do not need an online optimization in each step, which is often computational costly. Instead, \RL-agents try to find an optimal control policy during an offline training phase before they are implemented in real-world application \cite{Gorges.2017}. However, many modern \RL~algorithms are model-free and do not require model knowledge. Therefore, \RL~control methods can not only be trained in simulations but also in the field applications and optimize their control with respect to all the physical and parasitic effects as well as nonlinearities. Additionally, the same \RL~model architecture could be trained to control many different motors without expert's modification, similar to the \RL-agent that learns to play 
different Atari games\cite{Mnih.19.12.2013}.

The authors' contribution to this research field is the development of a toolbox for training and validation of \RL~motor controllers called gym-electric-motor (\gymem)\footnote{This package is available at \url{https://github.com/upb-lea/gym-electric-motor}}. It is based on \openai~environments \cite{Brockman.652016}.
Furthermore, different open-source \RL~toolboxes like Keras-rl \cite{Plappert.2016}, Tensorforce \cite{tensorforce} or OpenAI Baselines \cite{baselines} build upon the \openai~interface, which adds to its prevalence.
For easy and fast development, \RL-agents can be designed with those toolboxes and afterwards trained and  tested with \gymem~before applying them to real-world motor control. 

Currently, the \gymem~toolbox contains four different DC motors, namely the series motor, shunt motor, permanently excited motor and the externally excited motor as well as the three-phase permanent magnet synchronous motor (\PMSM).
In practical applications, power electronic converters are used in between the motor and a DC link to provide a variable input voltage.
Various converters provide different output voltage and current ranges, which affect the control behavior.
Therefore, different converters are included in the simulation as well as a mechanical load model. All models can be parametrized by the user.  To the authors' best knowledge, this is the first time an open-source toolbox for the development of \RL~electric motor controllers is published.

The paper is organized as follows:
A brief introduction into \RL~is given in Sec.~\ref{sec:RLsetting}, while the technical background on modelling electric drives for the control purpose is adressed in Sec.~\ref{sec:technicalbackground}. Then, details of the toolbox are presented in Sec.~\ref{sec:Toolbox}, followed by an example in Sec.~\ref{sec:example} with the comparison of a \DDPG-agent and a cascaded PI-controller resulting in a baseline for future research.
Finally, the paper is concluded and a research outlook is given in Sec.~\ref{sec:conclusion}. 

In this paper, variables that can be vector quantities are denoted in bold letters (e.g. $\bm{u}_{in}$) in any case whereas quantities that are always scalar are denoted in regular letters (e.g. $R_A$).

\section{Basic Reinforcement Learning Setting} 
\label{sec:RLsetting}
\begin{figure}
	\centering
	\includegraphics[width=0.6\columnwidth]{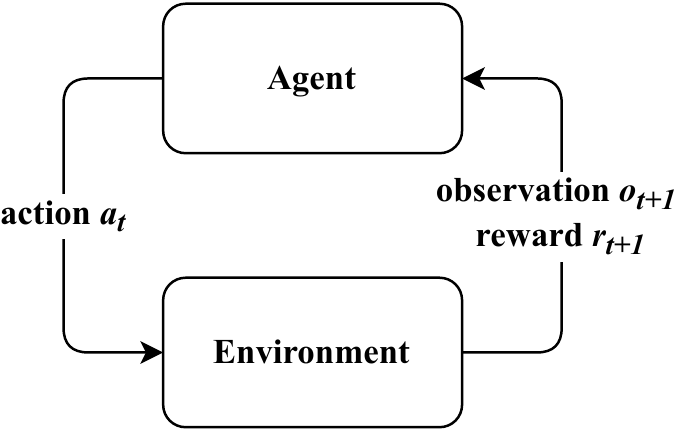}
	\caption{Basic reinforcement learning setting}
	\label{fig:RLsetting}
\end{figure}

A short introduction into \RL~is given, which shall clarify concepts and definitions for further reading. As depicted in Fig.~\ref{fig:RLsetting}, the basic RL setting consists of an agent and an environment. The environment can be seen as the problem setting and the agent as problem solver.
At every time step $t$, the agent performs an action $\action_t \in A$ on the environment. This action affects the environments state, which is updated based on the previous state $\state_t \in S$ and the action $\action_t$ to $\state_{t+1}$. Afterwards, the agent receives a reward $r_{t+1}$ for taking this action, and the environment shows the agent a new observation of the environment $\observ_{t+1}$. For example, in the motor control environments the observations are a concatenation of environment states and references. Based on the new observation, the agent will calculate a new action $\action_{t+1}$.

The goal of the agent is to find an optimal policy $ \pi: S \rightarrow A $. A policy $\pi$ is a function that maps the set of states $S$ to the set of actions $A$. An optimal policy maximizes the expected cumulative reward over time. Due to the dynamic character of the environment, the state and the reward at a timestep $t$ depend on many actions taken previously. Therefore, the reward for taking an action is often delayed over multiple time steps. A comprehensive introduction to \RL~is given in \cite{Sutton.2018}. 

In the case of motor control, the controller acts as agent and an environment includes the motor model and the reference trajectories. The agent receives a reward depending on how close the motor is following its reference trajectory.

\section{Technical Background}
\label{sec:technicalbackground}
\begin{figure}[!t]
	\centering 
	\includegraphics[width=0.8\columnwidth]{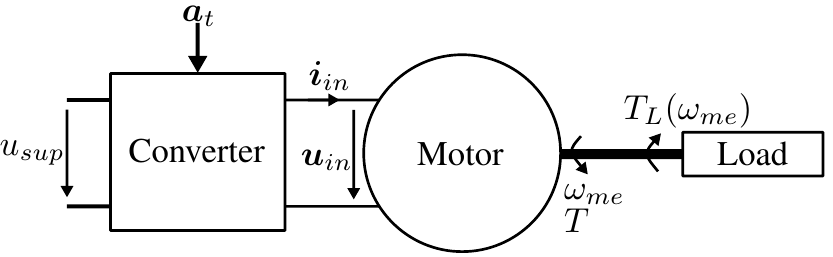}
	\caption{Scheme of converter, motor and load}
	\label{fig:Scheme}
\end{figure}

\gymem's environments simulate combinations of converter, electric motor and load, depicted in Fig. \ref{fig:Scheme}. This section includes short explanations of all included technical models.

\subsection{Basic Models of Electric Motors}
\label{sec:electricalmotors}
In general, electric motors can be represented by a system of differential equations (ODE system) in which the mechanical angular velocity and the currents are the motor states. All variables and motor constants are explained in Tab.~\ref{tab:variables}. The differential equation for the mechanical angular velocity 
\begin{equation}
\frac{\de \omega_{me}}{\de t}=\frac{T-T_L(\omega_{me})}{J}
\label{eq:omega}
\end{equation} holds for every rotary electric motor. A PMSM may have more than one pole pair, thus $p>1$. In this case the electrical angular velocity $\omega$ is $\omega=p \omega_{me}$. DC motors have only one pole pair and consequently, $\omega=\omega_{me}$ is valid. The mechanical angle $\varepsilon_{me}$, necessary for position control tasks, is given by \begin{equation}
	\frac{\de \varepsilon_{me}}{\de t}=\omega_{me}=\frac{1}{p}\frac{\de \varepsilon}{\de t}=\frac{\omega}{p}
\end{equation}
with the electrical angle $\varepsilon$.

All types of DC motors will be derived from the externally excited motor. The \PMSM~is different due to its three-phased feeding. Detailed explanations can be found in \cite{Bocker.08.01.2018, Bocker.11.07.2018, Chiasson.2005}.









\subsubsection*{Externally Excited Motor} 
\begin{figure}[!t]
\centering
\includegraphics[width=\columnwidth]{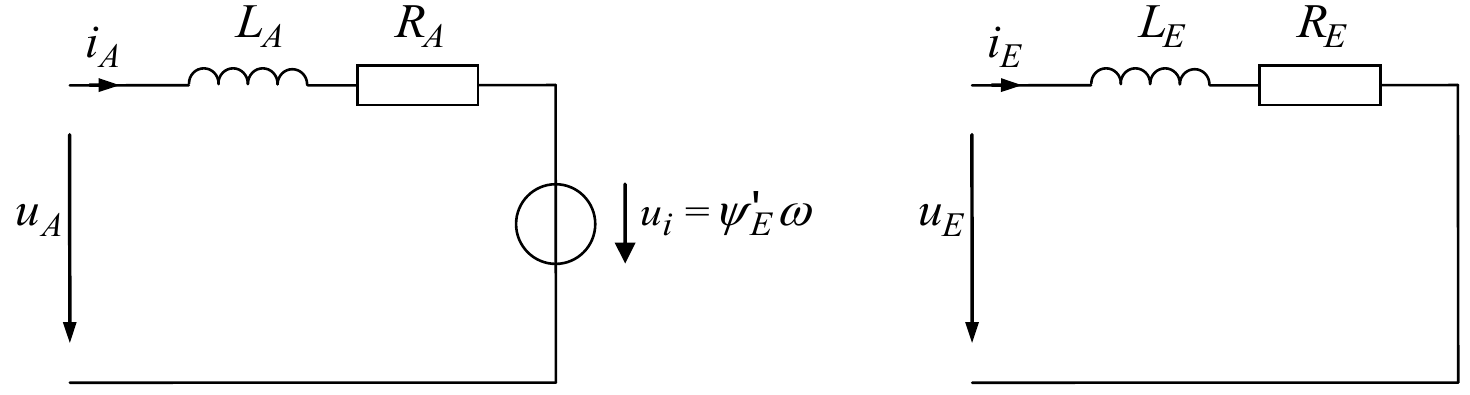}
\caption{Circuit diagram of externally excited motor (cf.~\cite{Bocker.08.01.2018})}
\label{fig:ExtEx}
\end{figure}

~ \\
The externally excited motor, as shown in Fig. \ref{fig:ExtEx}, consists of an armature and an excitation circuit with
\begin{align}
u_A&=\mathit{\Psi}^\prime_E \omega + L_A \frac{\de i_A}{\de t} +R_A i_A
\label{eq:u_a} \\
u_E&=L_E \frac{\de i_E}{\de t} + R_E i_E \text{.}
\label{eq:u_e}
\end{align}
The torque of the motor is given by
\begin{equation}
T=\mathit{\Psi}^\prime_E i_A
\label{eq:T}
\end{equation}
with the effective excitation flux
\begin{equation}
\mathit{\Psi}^\prime_E=L^\prime_E i_E\text{.}
\label{eq:psi_E}
\end{equation}
\eqref{eq:omega}-\eqref{eq:psi_E} form the following ODE system
\begin{equation}
\renewcommand{\arraystretch}{1.3}
		\begin{pmatrix}
	\frac{\de i_A}{\de t}\\ \frac{\de i_E}{\de t}\\ \frac{\de \omega}{\de t}
	\end{pmatrix}=\begin{pmatrix}
	\frac{1}{L_A}(u_A-L^\prime_E i_E \omega -R_A i_A)\\
	\frac{1}{L_E}(u_E-R_E i_E)\\
	\frac{1}{J}(L^\prime_E i_E i_A -T_L(\omega))
	\end{pmatrix}
	\label{eq:ExtExSystem}
\end{equation}
with the states $i_A$, $i_E$ and $\omega$. Further DC motor types can be derived with different combinations of armature and excitation circuits. Here, \eqref{eq:omega} to \eqref{eq:psi_E} hold for nearly all DC motors.

\subsubsection*{Shunt Motor} ~ \\
A shunt motor consists of a parallel connection of armature and excitation circuit. Hence, the voltages are the same ${\bm{u}_{in}=u=u_A=u_E}$ and the currents are summed up ${\bm{i}_{in}=i_A+i_E}$. However, the state is the same as for the externally excited motor and consists of $i_A$, $i_E$ and $\omega$, whereas the ODE system is
\begin{equation}
\renewcommand{\arraystretch}{1.3}
\begin{pmatrix}
\frac{\de i_A}{\de t}\\ \frac{\de i_E}{\de t}\\ \frac{\de \omega}{\de t}
\end{pmatrix}=\begin{pmatrix}
\frac{1}{L_A}(u-L^\prime_E i_E \omega -R_A i_A)\\
\frac{1}{L_E}(u-R_E i_E)\\
\frac{1}{J}(L^\prime_E i_E i_A -T_L(\omega))
\end{pmatrix}\text{.}
		\label{eq:ShuntSystem}
\end{equation}

\subsubsection*{Series Motor} ~ \\
As indicated by its name, the circuits are connected in series. Consequently, the armature and excitation currents are the same  $\bm{i}_{in}=i=i_A=i_E$ and the voltages are summed up to $\bm{u}_{in}=u=u_A+u_E$. The state contains $i$ and $\omega$ and the resulting ODE system is:
\begin{equation}
\renewcommand{\arraystretch}{1.3}
\begin{pmatrix}
\frac{\de i}{\de t}\\\frac{\de \omega}{\de t}
\end{pmatrix}=\begin{pmatrix}
\frac{1}{L_A+L_E}(-L^\prime_E i \omega - (R_A + R_E) i + u)\\
\frac{1}{J}(L^\prime_E i^2 -T_L(\omega))
\end{pmatrix}
\label{eq:SeriesSystem}
\end{equation}

\subsubsection*{Permanently Excited DC Motor} ~ \\
The permanently excited DC motor has permanent magnets for the excitation. Therefore, there is no excitation circuit but a constant excitation flux $\mathit{\Psi}^\prime_E$. The state of the motor consists of $i=i_A=\bm{i}_{in}$ and $\omega$, similar to the series motor. The ODE system reads
\begin{equation}
\renewcommand{\arraystretch}{1.3}
\begin{pmatrix}
\frac{\de i}{\de t}\\\frac{\de \omega}{\de t}
\end{pmatrix}=\begin{pmatrix}
\frac{1}{L_A} (-\mathit{\Psi}^\prime_E \omega - R_A i + u)\\
\frac{1}{J}(\mathit{\Psi}^\prime_E i - T_L(\omega))
\end{pmatrix}\text{.}
\label{eq:PermExSystem}
\end{equation}

\subsubsection*{Three-Phase Permanent Magnet Synchronous Motor} 
\begin{figure}
	\centering
	\includegraphics[width=0.8\columnwidth]{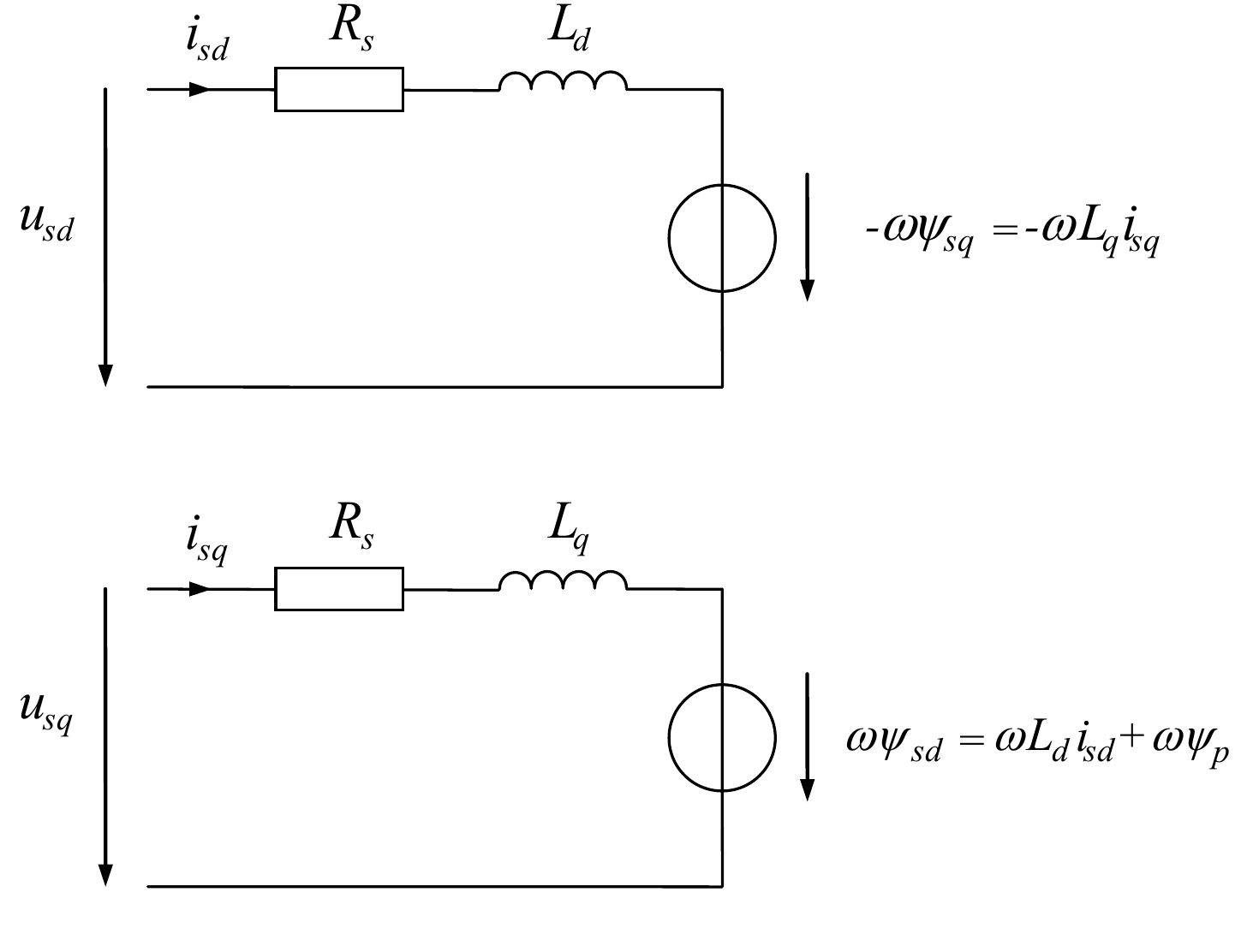}
	\caption{Circuit diagram of a \PMSM~in $d/q$-coordinates (cf.~\cite{Bocker.11.07.2018})}
	\label{fig:PMSM}
\end{figure}

~ \\
A \PMSM~consists of three phases with the phase voltages $u_a$, $u_b$ and $u_c$ and the phase currents $i_a$, $i_b$ and $i_c$. In order to simplify the mathematical representation two transformations are performed. First, the three quantities $x_a$, $x_b$ and $x_c$ are transformed with \eqref{eq:abctoalphabeta} to $x_\alpha$, $x_\beta$ and a zero component $x_0=0$. It is zero, because of the symmetric star connected PMSM without neutral conductor~\cite{Bocker.11.07.2018}. 
\iftrue
\begin{equation}
\begin{pmatrix}
x_\alpha\\x_\beta\\x_0
\end{pmatrix}=\begin{pmatrix}
\frac{2}{3}&-\frac{1}{3}&-\frac{1}{3}\\
0&\frac{1}{\sqrt{3}}&-\frac{1}{\sqrt{3}}\\
\frac{\sqrt{2}}{3}&	\frac{\sqrt{2}}{3}&	\frac{\sqrt{2}}{3}
\end{pmatrix}\begin{pmatrix}
x_a\\x_b\\x_c
\end{pmatrix}
\label{eq:abctoalphabeta}
\end{equation}
Second, the quantities are transformed to rotor fixed coordinates $d$ and $q$ using the angle of the rotor flux $\varepsilon$ and the transformation matrix
\begin{equation}\begin{pmatrix}
x_d\\x_q
\end{pmatrix}=
\begin{pmatrix}
\cos(\varepsilon)&\sin(\varepsilon)\\
-\sin(\varepsilon)&\cos(\varepsilon)
\end{pmatrix}	\begin{pmatrix}
x_\alpha\\x_\beta
\end{pmatrix}\text{.}
\label{eq:alphabetatodq}
\end{equation}
\fi
A similar reverse transformation to the $a,b,c$ domain is possible as given in~\cite{Bocker.11.07.2018}.
After transformations \eqref{eq:abctoalphabeta} and \eqref{eq:alphabetatodq}, the circuits result in
\begin{align}
\centering
u_{sd}&=R_s i_{sd}+L_d \frac{\de i_{sd}}{\de t}-\omega_{me}p L_q i_{sq}\\
u_{sq}&=R_s i_{sq}+L_q \frac{\de i_{sq}}{\de t}+\omega_{me}p L_d i_{sd}+\omega_{me}p \mathit{\Psi}_p	
\end{align}
as shown in \figurename~\ref{fig:PMSM}.
The torque equation reads
\begin{equation}
T=\frac{3}{2}p(\Psi_p + (L_d - L_q) i_{sd}) i_{sq}
\end{equation}
and the angular velocity is also given by \eqref{eq:omega}. Hence, the ODE system
\begin{equation}
\renewcommand{\arraystretch}{1.3}
\begin{pmatrix}	
	\frac{\de i_{sd}}{\de t}\\
	\frac{\de i_{sq}}{\de t}\\
	\frac{\de \omega_{me}}{\de t}\\
	\frac{\de \varepsilon_{me}}{\de t}
\end{pmatrix}
=
\begin{pmatrix}
	\frac{1}{L_d}(u_{sd}-R_s i_{sd}+L_q \omega_{me}p i_{sq})\\
	\frac{1}{L_q}(u_{sq}-R_s i_{sq}-\omega_{me}p (L_d i_{sd}+\mathit{\Psi}_p))\\
	\frac{1}{J}(T-T_L(\omega_{me}))\\
	\omega_{me}
\end{pmatrix}
\label{eq:PMSMSystem}
\end{equation}
consists of the states $i_{sd}$, $i_{sq}$, $\omega_{me}$ and $\varepsilon_{me}$.
\subsection{Basic Models of Power Electronic Converters}
\label{sec:converter}
\begin{figure}[!t]
	\centering
	\begin{subfigure}[b]{0.45\columnwidth}
		\raggedbottom
			\includegraphics[width=\columnwidth, trim=0cm 0cm 4.5cm 0cm, clip=true]{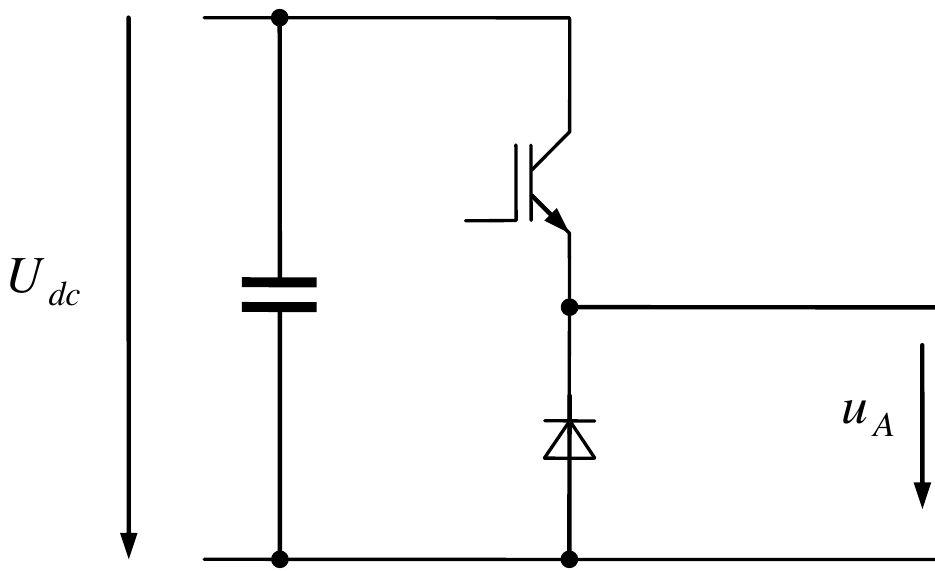}
			\caption{1QC}
	\end{subfigure}
	\begin{subfigure}[b]{0.45\columnwidth}
	\includegraphics[width=\columnwidth]{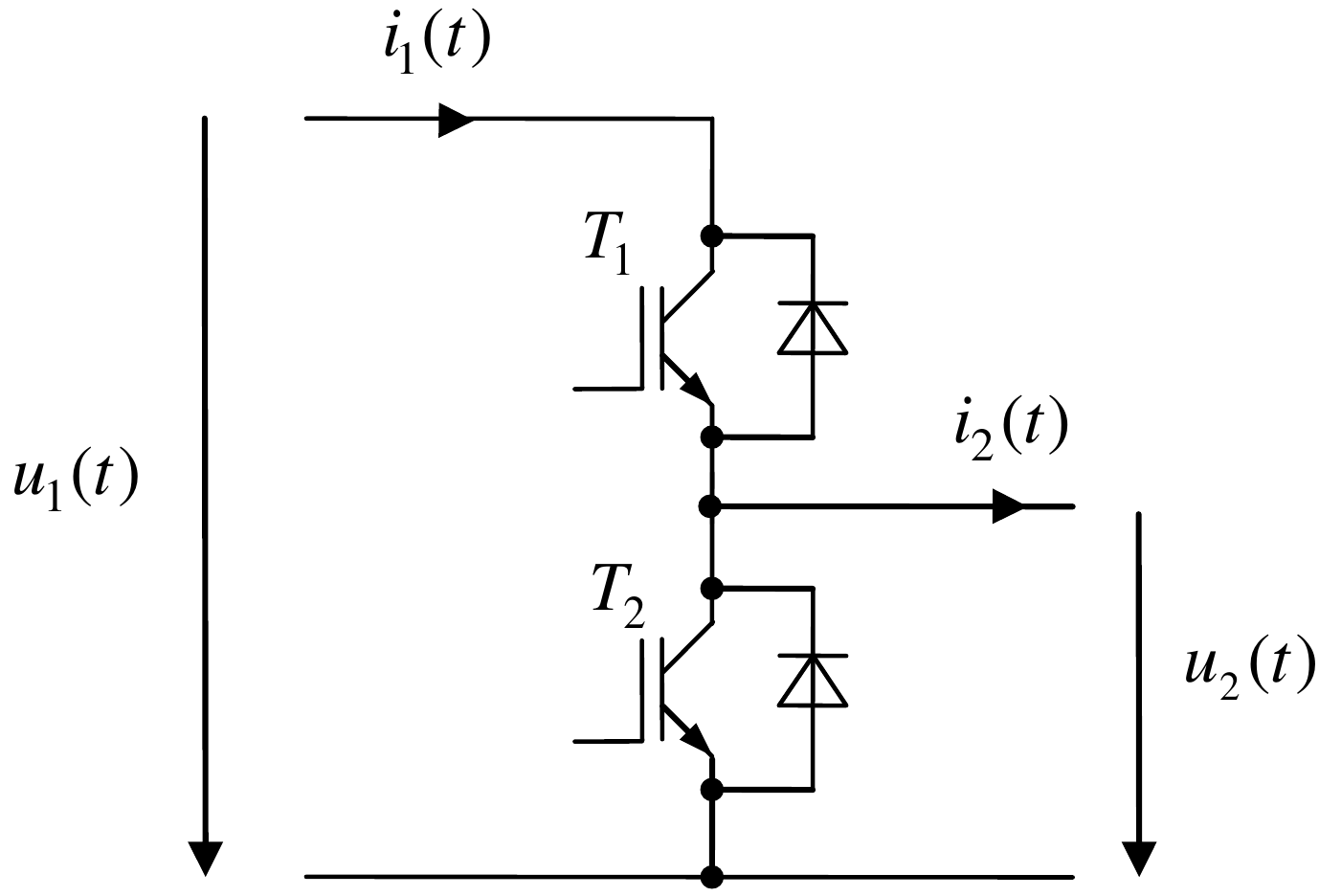}
	\caption{2QC}
		\end{subfigure}
\begin{subfigure}[b]{0.45\columnwidth}
	\includegraphics[width=\columnwidth, , trim=0cm 0cm 3.5cm 0cm, clip=true]{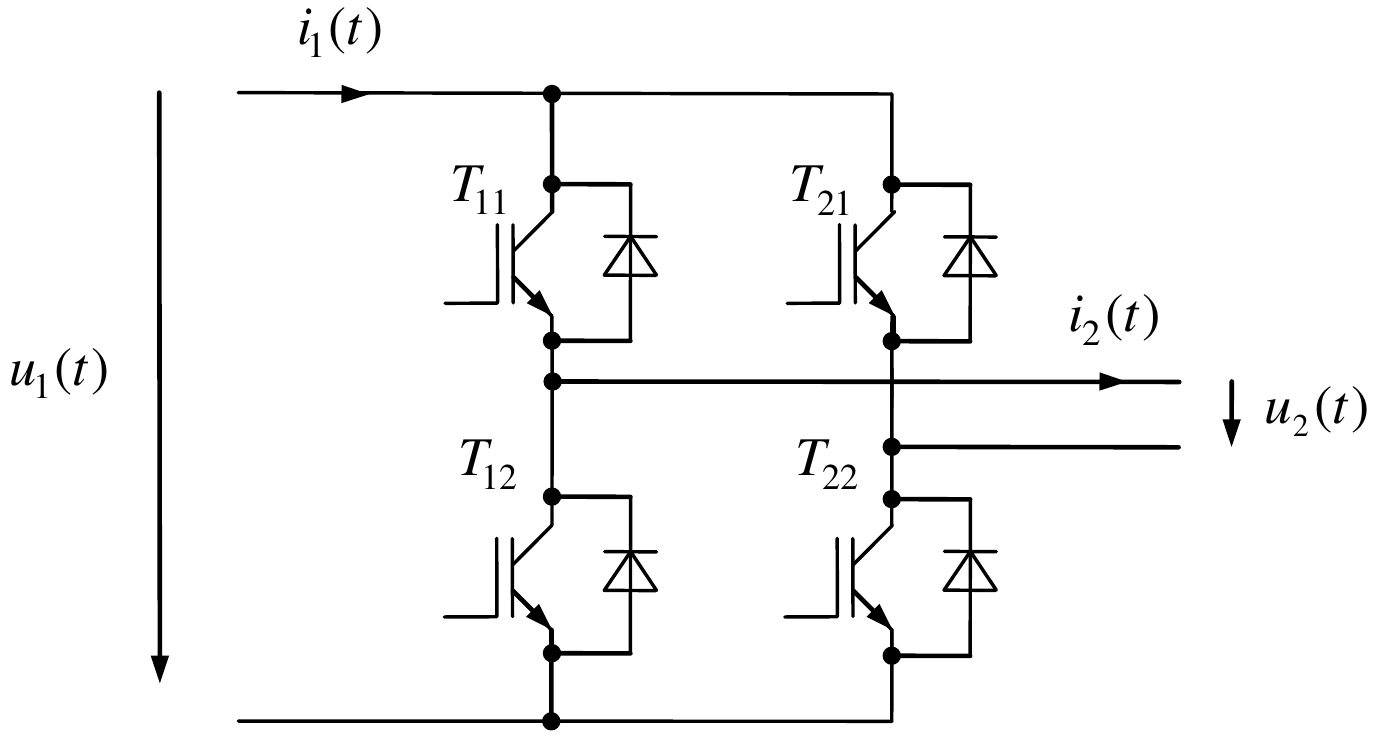}
	\caption{4QC}
		\end{subfigure}
\begin{subfigure}[b]{0.45\columnwidth}
	\includegraphics[width=\columnwidth,, trim=0cm 0cm 1.5cm 0cm, clip=true]{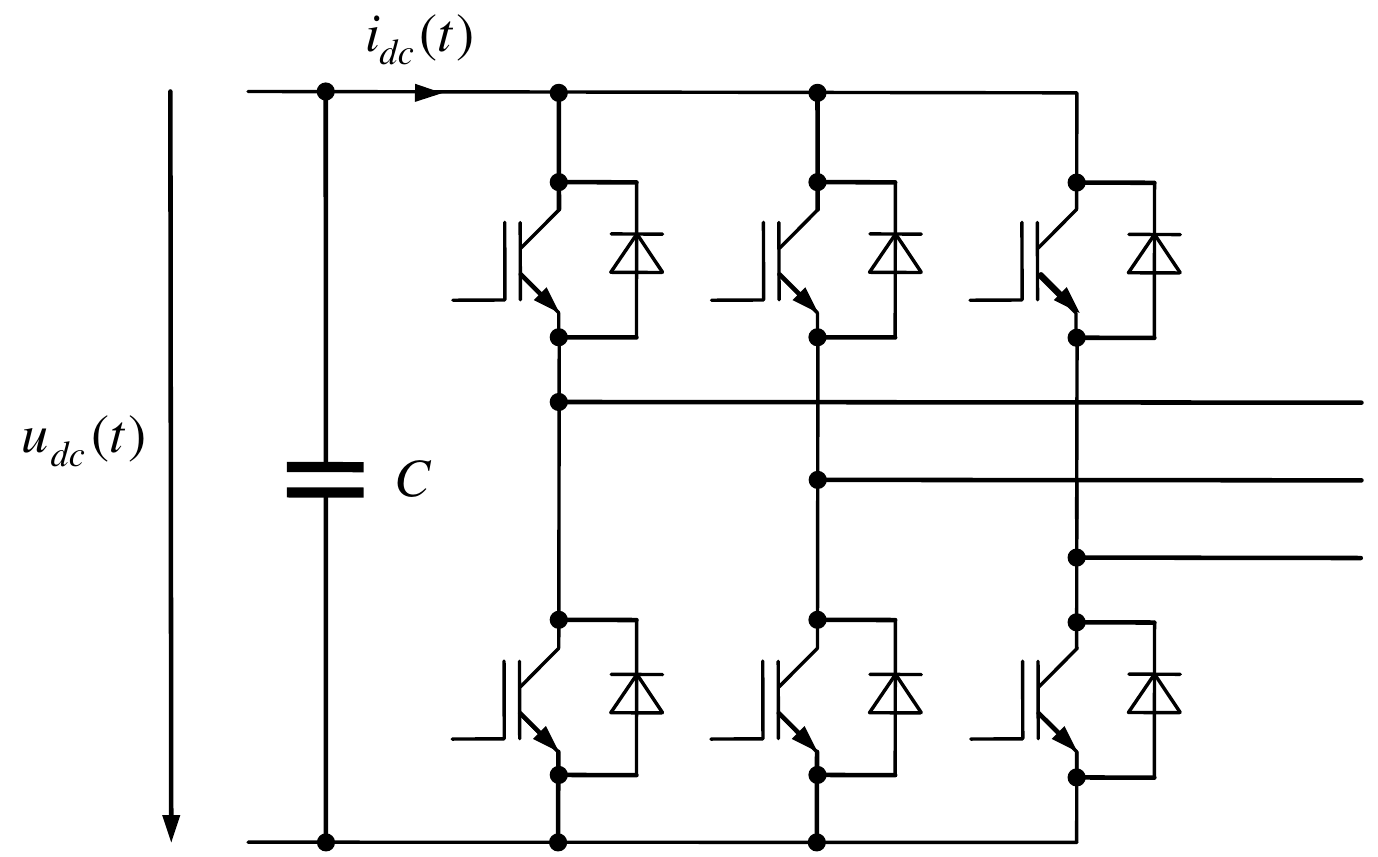}
	\caption{three-phase B6 bridge}
		\end{subfigure}
	\caption{Different converter topologies (cf.~\cite{Bocker.08.01.2018, Bocker.11.07.2018})}
	\label{fig:Converter}
\end{figure}

In practical applications, the motor often shall run at different velocities and, thus, the input voltage is not constant. To achieve variable input voltages, a power electronic converter is used in between the electric motor and the DC link (i.e. the supply voltage which could be a battery or a rectified grid supply). The following DC converters, depicted in \figurename~\ref{fig:Converter}, are covered in the \gymem~toolbox for feeding the DC motors ~\cite{Bocker.08.01.2018}:
\begin{itemize}
	\item 1 quadrant converter (1QC), also called buck converter
	\item 2 quadrant converter (2QC) as asymmetric half bridge
	\item 4 quadrant converter (4QC).
\end{itemize}
For the thre-phase \PMSM~a
\begin{itemize}
	\item B6 bridge three-phase converter
\end{itemize}
is implemented~\cite{Bocker.11.07.2018}.
A B6 bridge can be seen as three parallel 2QC, so one 2QC for each phase. Power electronic converters are switched systems, thus, different switching schemes determine the resulting three-phased voltage.

The inputs of a converter are the supply voltage and direct switching commands for the transistors to switch them on or off. Typical controllers provide either a desired output voltage or a duty cylce in a normalized form.
This continuous value needs to be mapped to a switching pattern over time.
Common approaches are pulse width or space-vector modulation (PWM/SVM). Then, the average output voltage over one pulse period of the power electronic converter equates the requested voltage. From simulation point of view, this would require very tiny time steps to cover the switching instants accurately. However, to speed up the simulation the modulation schemes are neglected and a dynamic average model is used \cite{Prof.Dr.JoachimBocker.Sommersemester2019}.

Moreover, a dead time of one sampling time step and a user-parametrized interlocking time can be considered in all converters to account for these common delays in real applications. 
The ranges of the normalized output voltages and currents as well as the possible switching states are presented in Tab. \ref{tb:DCconverter}.

\begin{table}[h]
	\renewcommand{\arraystretch}{1.3}
	\centering
	\caption{Possible voltage and current ranges of the converter and the number of switching states.}
	\label{tb:DCconverter}
	\begin{tabular}{c|ccccc}
		\toprule
		&$u\geq0$&$u<0$&$i\geq 0$&$i<0$&\#switching states\\
		\midrule
		1QC & x &  & x &  & 2 \\ \hline
		2QC & x &  & x & x & 3 \\ \hline
		4QC & x & x & x & x & 4 \\ \hline
		B6 & x & x & x & x & 8 \\ 
		\bottomrule
	\end{tabular}
\end{table}


\subsection{Basic Model of the Load}
The attached mechanical load in the toolbox is represented by the function
\begin{equation}
T_L(\omega)= \mathrm{sign}(\omega_{me}) (c \omega^2_{me} + \mathrm{sign}(\omega_{me}) b \omega_{me} + a)
\label{eq:load}
\end{equation} with a constant load torque $a$, viscous friction coefficient $b$ and aerodynamic load torque coefficient $c$. These parameters as well as a moment of inertia of the load $ J_{load}$ can be freely defined by the user to simulate different loads.

\subsection{Discretization}
The \RL-agents act in discrete time and, therefore, the continuous-time ODE systems \eqref{eq:ExtExSystem}, \eqref{eq:ShuntSystem}, \eqref{eq:SeriesSystem}, \eqref{eq:PermExSystem} and \eqref{eq:PMSMSystem} need to be discretized for the simulation. Standard methods as Euler's method or Runge-Kutta method \cite{Butcher.2008} can be applied for discretization in the toolbox.

\section{The Gym Electric Motor Toolbox}
\label{sec:Toolbox}
\begin{figure}[!t]
	\centering
	\includegraphics[width=\columnwidth]{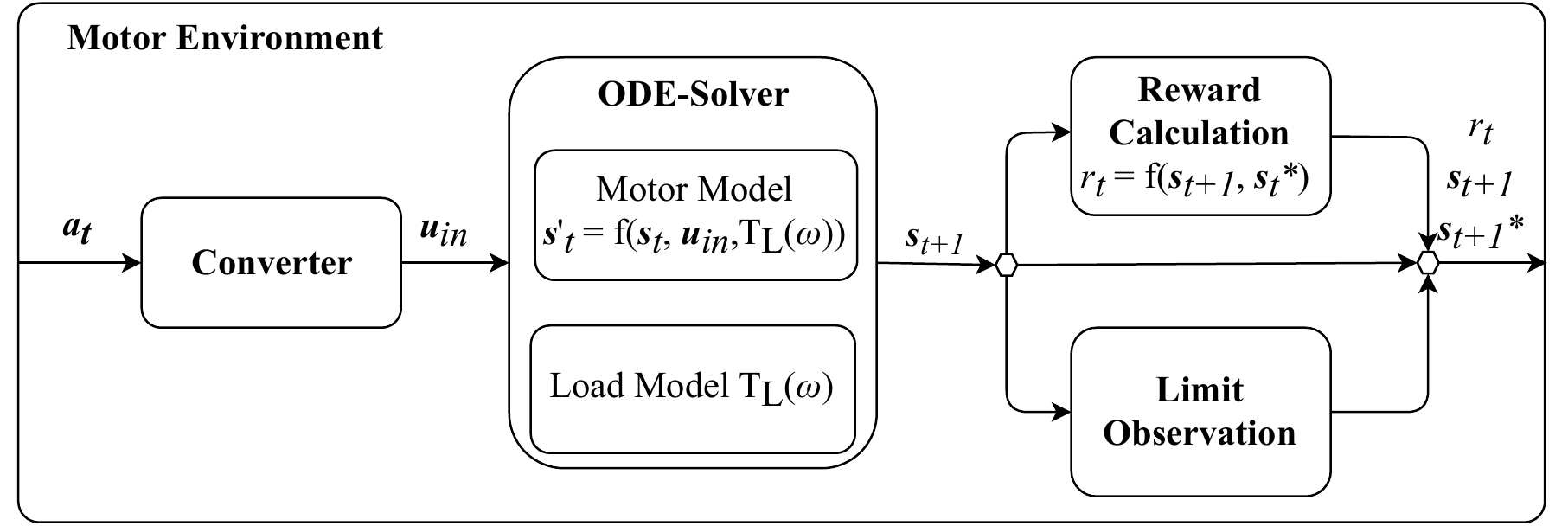}
	\caption{Control flow from an action to a new observation}
	\label{fig:EnvStep}
\end{figure}
In this section, the main structure, key features and the interface of the electric motor environments  are presented.

Each motor environment belongs to one specific pair of motor and action type (see Sec.~\ref{subsec:actionspace}). The user specifies the control purpose, e.g. speed or torque control by selecting different reward weights. Furthermore, it can be specified, how many future reference points are revealed to the agent, by selecting the prediction horizon.

The general simulation setup is episode-based like many other RL problems and Gym environments. This means that the environment has to be reset before a new episode starts and it performs cycles of actions and observations like it is shown in Fig.~\ref{fig:RLsetting}. An episode ends, when a safety limit of the motor is violated or a maximum number of steps has been performed. Then, the motor environment is reset to a random initial state and new references for the next episode are generated.

The control flow during one motor environment simulation step is shown in \figurename~\ref{fig:EnvStep}.
The control action $\bm{a}_t$ is converted to an input voltage $\bm{u}_{in}$ of the motor.
Then, the next state $s_{t+1}$ is calculated using an ODE solver. This solver uses the motors differential equations including the load torque \eqref{eq:load}. Afterwards, the reward $r_{t+1}$ is calculated based on the current state and current reference $s^*_{t+1}$. If a state exceeds the specified safety limits, the limit observer stops the episode and the lowest possible reward is returned to the agent to punish the limit violation.
The user can specify which states are visualized in graphs as depicted in \figurename~\ref{fig:ExampleTrajectory}.

\subsection{Action Space}
\label{subsec:actionspace}
In an \openai~environment the action- and observation space define the set of possible values for the actions and observations.
In terms of electric motor control, the action space could be modeled in a discrete or continuous way.
\subsubsection*{Continuous Action} In the continuous case the action is the desired duty cycle that should be utilized in the converter and its range is the same as the normalized voltage range as given in Tab.~\ref{tb:DCconverter}. Consequently, a PI-controller or a \DDPG-agent can be used for a motor control of this type.
From a MPC point of view, this case is known as  a continuous control-set (CCS).
\subsubsection*{Discrete Action} In the discrete case, the actions are the direct switching commands for the transistors. Potential controllers are a hysteresis on-off controller or a \DQN-agent.
From an MPC point of view, this is known as a finite control-set (FCS).

\subsection{Observation Space}
The observations of the environments are a concatenation of the environment state and the reference values the controller should track. All values are normalized by the limits of the state variables to a range of $[-1,+1]$ or $[0,+1]$ in case negative values are implausible for a state. The environment state is its motor state extended by the torque and the input voltages, as given for each motor type:

\begin{subequations}
	\label{eq:envstates}
	\begin{align}
	s_{ExtEx} &= [ \omega, T, i_A, i_E, u_A, u_E, u_{sup}] \\
	s_{Shunt} &= [ \omega, T, i_A, i_E, u, u_{sup}] \\
	s_{Series}&= [ \omega, T, i, u, u_{sup}] \\
	s_{PermEx}&= [ \omega, T, i, u, u_{sup}] \\
	s_{PMSM} &= [ \omega, T, i_a, i_b, i_c, u_a, u_b, u_c, u_{sup}, \varepsilon]
	\end{align}
\end{subequations}

For example, each observation for a PMSM environment for current control and a prediction horizon of two (current and next reference value presented) looks as follows:
\begin{equation}
o_t = [\omega,\dots, u_{sup}, i^*_{a,t}, i^*_{a, t+1}, i^*_{b,t}, i^*_{b, t+1}, i^*_{c,t}, i^*_{c, t+1}]
\end{equation}

\subsection{Rewards}
\label{subsec:rewards}
Different reward functions and weights can be chosen. First, the user can specify a reward weight $w_{\{k\}}$ for each observation quantity $k$ of the $N$ environment state variables. Those should sum up to $1$ to receive rewards in the range of $[-1,~0]$ or $[0,+1]$, depending on the reward function. The reward weights specify which state reference quantities the agent should follow, because those are responsible for the reward the agent tries to maximize. Hence, environment states with $w_{\{k\}}=0$ have no tracked reference. A negative weighted sum of absolute (WSAE) and squared (WSSE) errors are available as reward functions, which result in larger negative reward the larger the difference between reference and state values is.
Furthermore, both reward functions are implemented in a shifted way ((SWSAE) and (SWSSE)) where the reward is incremented, such that perfect actions result in a reward of one, because some RL-agents' learning behavior is different for positive and negative rewards.
All reward functions are given below:
\subsubsection*{weighted sum of absolute error (WSAE)} 
\begin{equation}
r_t=-\sum_{k=0}^{N} w_{\{k\}} \vert s_{{\{k\}}t}-s^*_{{\{k\}}t} \vert
\end{equation}
\subsubsection*{weighted sum of squared error (WSSE)}
\begin{equation}
	r_t=-\sum_{k=0}^{N} w_{\{k\}} ( s_{{\{k\}}t}-s^*_{{\{k\}}t} )^2
\end{equation}
\subsubsection*{shifted weighted sum of absolute error (SWSAE)}
\begin{equation}
	r_t=1-\sum_{k=0}^{N} w_{\{k\}} \vert s_{{\{k\}}t}-s^*_{{\{k\}}t} \vert
\end{equation}
\subsubsection*{shifted weighted sum of squared error (SWSSE)}
\begin{equation}
r_t=1-\sum_{k=0}^{N} w_{\{k\}} ( s_{{\{k\}}t}-s^*_{{\{k\}}t} )^2
\end{equation}

\subsection{Limit Observation and Safety Constraints}
The typical operation range of electric motors is limited by the nominal values of each variable. However, the technical limits of the electric motor are larger. Those limits must not be exceeded to prevent motor damage, which might be inflicted due to excessive heat generation. Motors are stopped if limits are violated in real applications. The user can specify the nominal values and safety margin $\xi$. In the toolbox, the limits are determined as follows
\begin{equation}
	x_{limit} = \xi x_N \text{.}
\end{equation}
An important task for the control is to hold those limits.
Consequently, learning episodes will be terminated if limits are violated as in real applications, and a penalty term can be chosen that is affecting the final reward to account for those cases. The penalty can be a constant negative term or zero.
If the internal reward function returns positive rewards, then a zero reward penalty for violating limits is sufficient, because it is the worst the agent could get.
In case of negative rewards, if the penalty is too low, the agent could try to end the episodes by violating limits to maximize the cumulative reward.
To avoid this, a penalty term which is based on the Q-function \cite{Sutton.2018} can be selected. This term ensures that for every limit violating state the expected reward is lower than for non limit violating states, in case the $\gamma$ parameter is chosen equivalent to the RL-agents discount factor $\gamma$. The penalty term is  
\begin{equation}
r_t = -\frac{1}{1-\gamma} \text{.}
\end{equation}

\subsection{Reference Generation}
The generation of reference trajectories (e.g. the control set points) is a fundamental part of the environment and necessary for diverse training. The references should cover all use cases such that the \RL-agent generalizes well and to avoid biased training data. In order to achieve this, standard reference shapes are implemented, e.g. sinusoidal, asymmetric triangular, rectangular and sawtooth signals as depicted in \figurename~\ref{fig:AvailableReferences} with random time periods, amplitudes and offsets. Also pseudo-random references are available with respect to the limits and dynamics of the motors. Such a random reference for the angular velocity is used in \figurename~\ref{fig:ExampleTrajectory}. To achieve this, a random discrete fourier spectrum with limited bandwidth for the input voltage is generated for a whole episode. Afterwards, it is transformed to the time-domain and the inputs are applied to the motor, and all states are saved as the reference. To hold the limits, the references for each quantity are clipped to their nominal values to keep a safety margin.
For each new episode the shape of the reference is sampled from a uniform distribution. Each standard shape has a probability of $12.5~\%$ while random references appear half the time.
Furthermore, zero references for some states can be considered, for example if the input voltage or current should be minimized in order to reduce the power dissipation.
\iftrue
\begin{figure}[!t]
	\centering
	\begin{subfigure}[t]{0.22\columnwidth}
		\includegraphics[trim=1cm 1cm 1cm 1cm, clip ,width=\columnwidth]{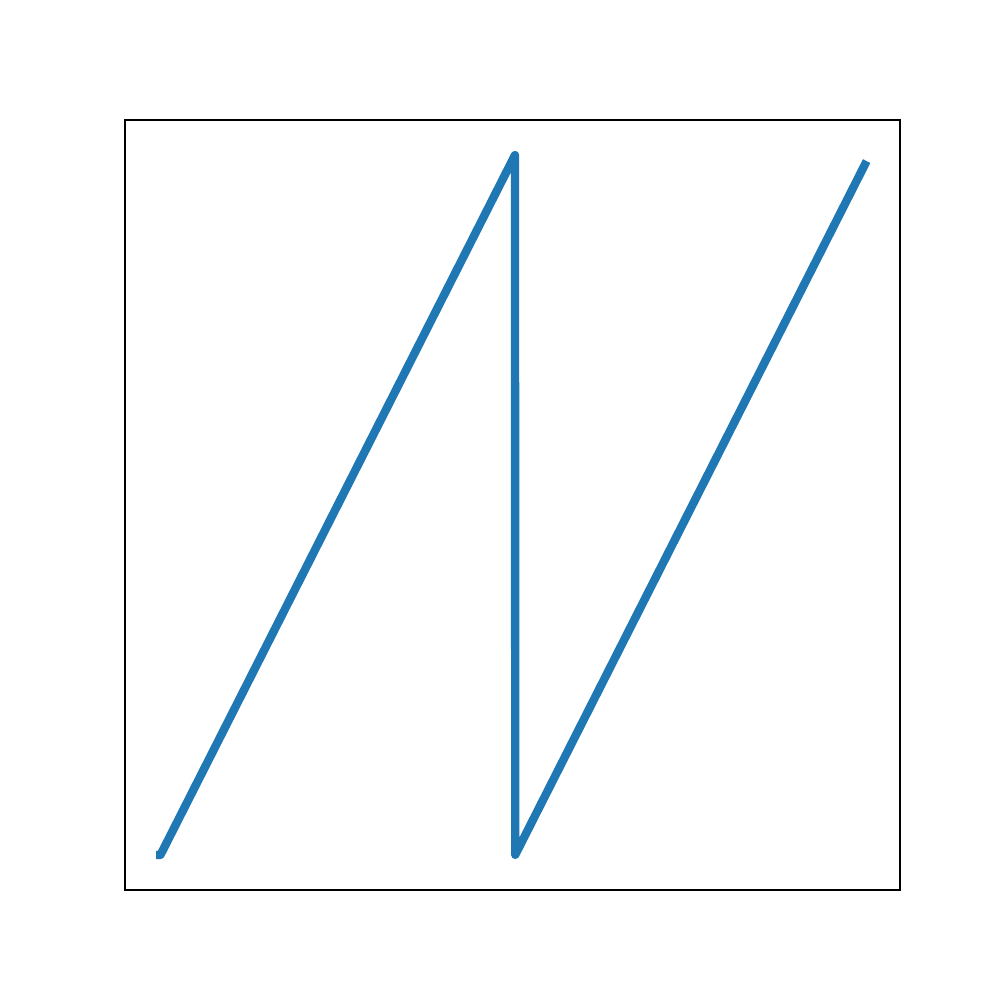}
		\caption{sawtooth}
	\end{subfigure}
	\begin{subfigure}[t]{0.22\columnwidth}
		\includegraphics[trim=1cm 1cm 1cm 1cm, clip ,width=\columnwidth]{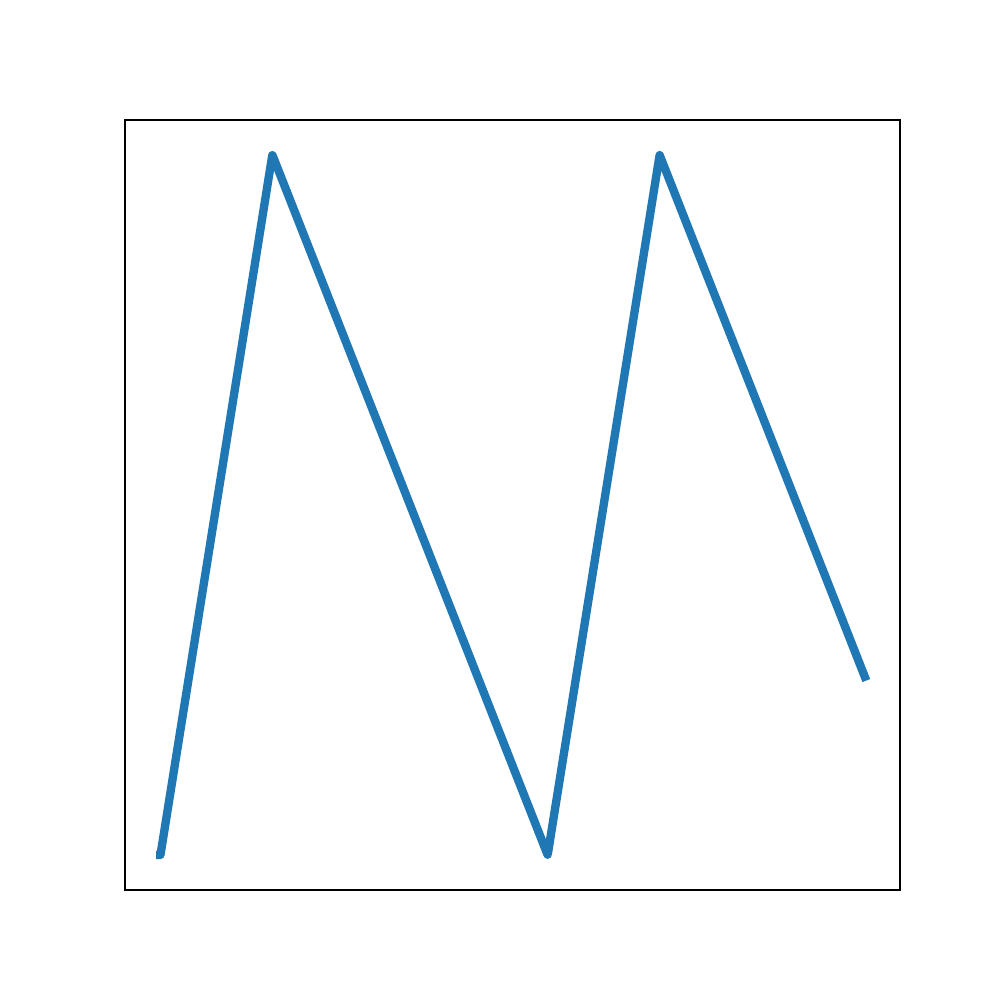}
		\caption{asymmetric triangular}
	\end{subfigure}
	\begin{subfigure}[t]{0.22\columnwidth}
	\includegraphics[trim=1cm 1cm 1cm 1cm, clip ,width=\columnwidth]{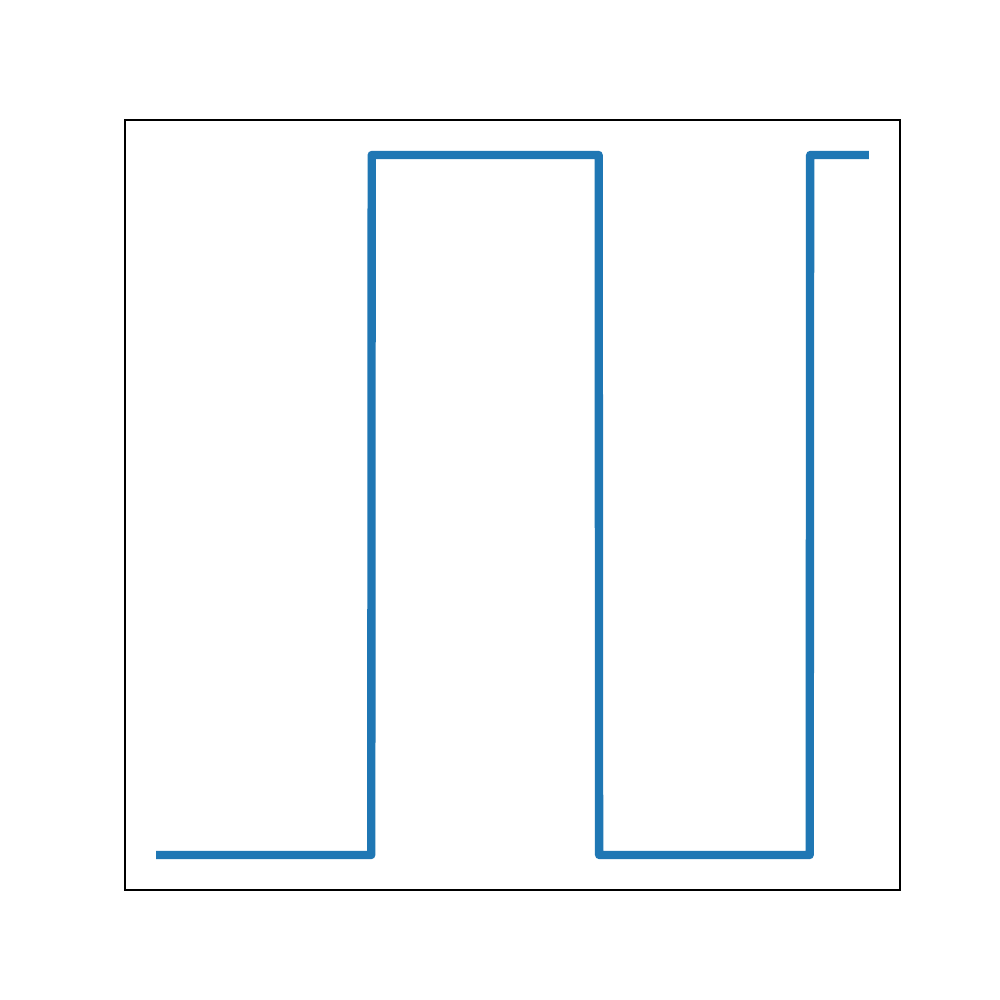}
	\caption{rectangular}
\end{subfigure}
\begin{subfigure}[t]{0.22\columnwidth}
	\includegraphics[trim=1cm 1cm 1cm 1cm, clip ,width=\columnwidth]{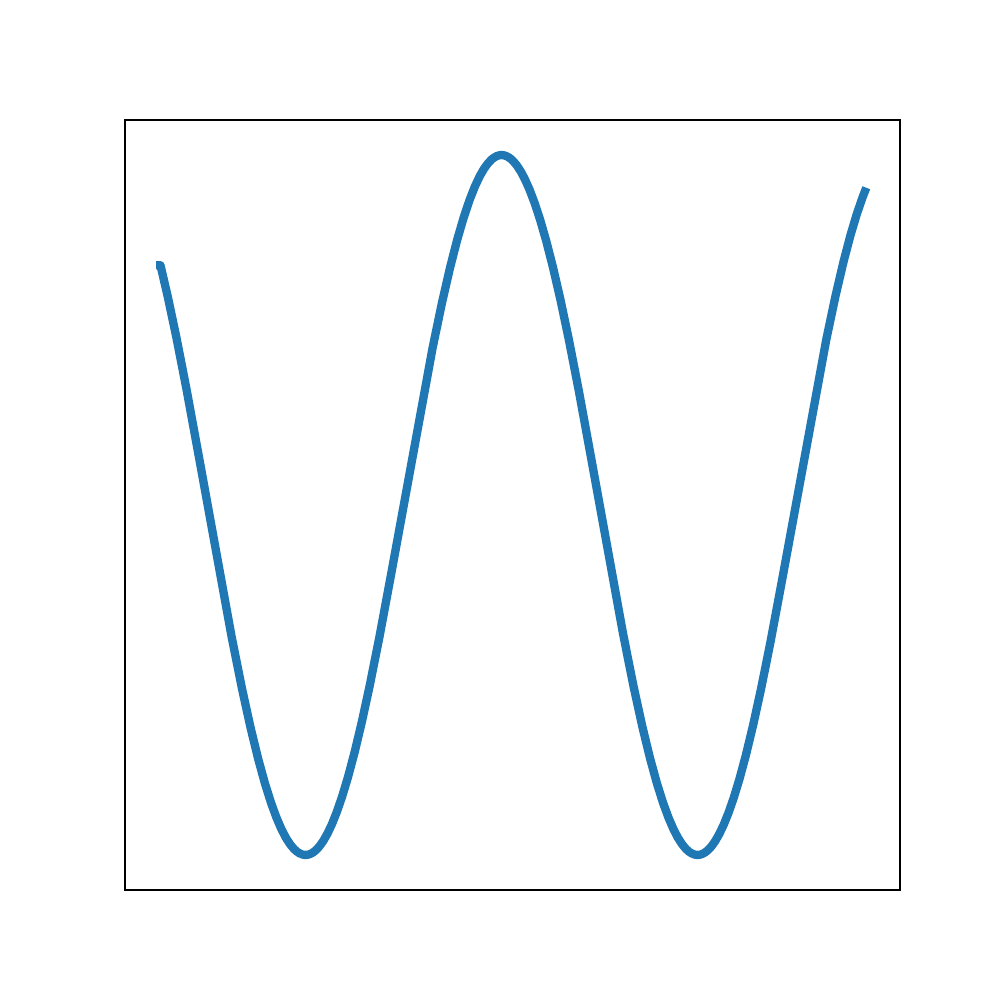}
	\caption{sinusoidal}
\end{subfigure}
	\caption{Available standard reference trajectory shapes}
	\label{fig:AvailableReferences}
\end{figure}
\fi
\subsection{Noise}
No real-world application is noise-free. In the toolbox, an additive white Gaussian distributed noise $\noise_k$ can be applied to the environment states. For each environment state $s_{k}$ a certain noise level $\rho_{k}$ can be selected. The noise level is defined as the ratio between noise power and signal power of the state. For a rough estimation of the signal power, each amplitude of the environment states was assumed to be distributed triangular between zero and its nominal value with its mode at zero. Therefore, the normalized noise added to each environment state is calculated as follows:
 \begin{equation}
 	\label{eq:noise}
 	\noise_{k}=\mathcal{N}(0, \frac{\rho_{k}}{6} \frac{1}{\xi^2})
\end{equation} 
The noise is modeled as measurement noise for each state which has only an effect on the observations except for the noise added to the input voltages of the motors (inverter nonlinearity). This is interpreted as input noise to the system.

\section{Example}
\label{sec:example}
In the following, an application example for the \gymem~toolbox is provided.
This example is to demonstrate the possibilities of the toolbox and the \RL~motor control approach.
First, the training and test setting is presented.
Afterwards, the training process of the agent is illustrated and then, the trained \DDPG-agent is compared with a cascaded PI-controller, where current is controlled in an inner loop and motor speed in an outer loop.
The PI-controller parameters are chosen as suggested in \cite{.2009}.

\subsection{Setting}
In this example, the toolbox is used to train a \DDPG-agent from Keras-rl with an actor and critic architecture as described in Tab.~\ref{tab:exampleactorparameter}.
The agent learns to control the angular velocity of a series DC motor with a continuous action space supplied by a 1QC.
Motor and load parameters are compiled in Tab.~\ref{tab:examplemotorparameter}.
The reward function is the SWSAE with reward weight $1$ on the angular velocity $\omega$ and $0$ otherwise.
The training consists of $7\,500\,000$ simulation steps partitioned in episodes of length $10\,000$.
Furthermore, a white Gaussian process is considered in the training algorithm to ensure exploratory behaviour to find the optimal control policy.
The power of the Gaussian process is decreased during training.
The equivalent real time of the simulation translates to \SI{12.5}{\minute}.

\begin{table}[htp]
	\renewcommand{\arraystretch}{1.3}
	\caption{Exemplary hyperparameters of an actor (left) and critic (right) network.}
	\label{tab:exampleactorparameter}
	\begin{center}
		\begin{tabular}{ccc}
		 \toprule
\textbf{Layer} & \textbf{Width} & \textbf{Activation} \\
\midrule
Input & 6 & / \\\hline
Dense & 64 & ReLU \\\hline
Dense & 1 & sigmoid \\
			\bottomrule
		\end{tabular}\hspace{6mm}
		\begin{tabular}{ccc}
				 \toprule
		\textbf{Layer} & \textbf{Width} & \textbf{Activation} \\
		\midrule
		Input & 7 & / \\\hline
		Dense & 64 & ReLU \\\hline
		Dense & 1 & linear \\
		\bottomrule
		\end{tabular}
	\end{center}
\end{table}

\begin{table}[h]
	\renewcommand{\arraystretch}{1.3}
	\caption{Example's motor and load parameter}
	\label{tab:examplemotorparameter}
	\begin{center}	
		\begin{tabular}{cc}
			\toprule
			\textbf{variable}&\textbf{value}\\ 
			\midrule
			$\tau$&\SI{1E-4}{\second}\\ \hline
			$R_A$&\SI{2.78}{\ohm} \\\hline
			$R_E$&\SI{1.0}{\ohm} \\\hline
			$L_A$&\SI{6.3}{\milli\henry}\\\hline
			$L_E$&\SI{1.6}{\milli\henry}\\\hline
			$L^\prime_E$&\SI{0.5}{\milli\henry}\\\hline
			$J_{rotor}$& \SI{17}{\gram/\meter^2}\\\hline
			$\omega_{N}$&\SI{368}{\radian/\second}\\
			\bottomrule
		\end{tabular}
		\hspace{8mm}
		\begin{tabular}{cc}
			\toprule
			\textbf{variable}&\textbf{value}\\
			\midrule	
			$T_{N}$&\SI{250}{\newton\meter} \\\hline
			$i_{N}$&\SI{50}{\ampere}\\\hline
			$u_{sup}$&\SI{420}{\volt}\\\hline
			$a$&\SI{0.01}{\newton\meter}\\ \hline
			$b$&\SI{0.12}{\newton\meter/\second}\\ \hline
			$c$&\SI{0.1}{\newton\meter/\second^2}\\ \hline
			$J_{load}$&\SI{1}{\kilogram/\meter^2}\\
			\bottomrule
		\end{tabular}
	\end{center}
\end{table}

\subsection{Results}

\begin{figure}[t]
	\centering
	\includegraphics[width=\columnwidth, trim=0cm 0.2cm 0cm 1.3cm, clip=true]{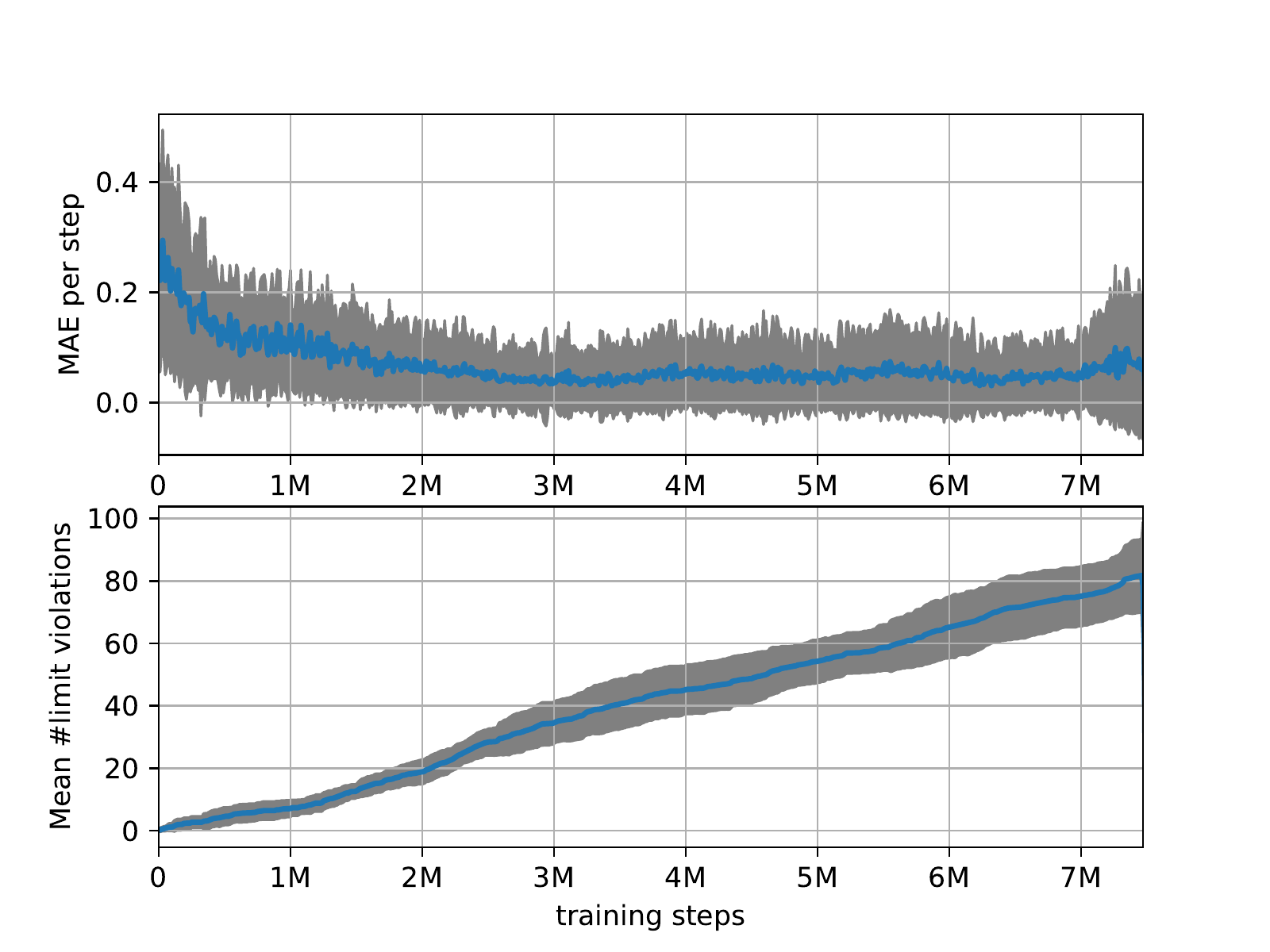}
	\caption{At the top, the \MAE~per training step of $10$ \DDPG-agents are presented. At the bottom the mean cumulative number of limit violations is plotted. The gray regions visualize the area inside the standard deviations.}
	\label{fig:LearningCurveRLAgent}
\end{figure}

The training process is depicted in \figurename~\ref{fig:LearningCurveRLAgent}. At the beginning, the \MAE~is $0.25$ and decreases to $0.04$ at $3\,000\,000$ steps. At the end, the \MAE~is $0.059$. The standard deviation decreases from $0.185$ at the beginning to $0.05$ and increases at the end, too.
In the bottom plot the mean cumulative number of limit violations during the training of $10$ \DDPG-agents is presented.
It increases approximately linearly, which means that the agent violates limits at the end of the training as frequently as at the beginning.
Reasons for this could be the Gaussian noise added to the control actions.
The agent tries to set the quantities (e.g. current) to their maximum allowed value for optimal control.
Then, little noise is sufficient to exceed the limit.
Furthermore, the limit violations show, that the agent does not learn to hold the limits.



\begin{figure}[!t]
	\includegraphics[width=\columnwidth, trim=0cm 0.3cm 1cm 1.4cm, clip=true]{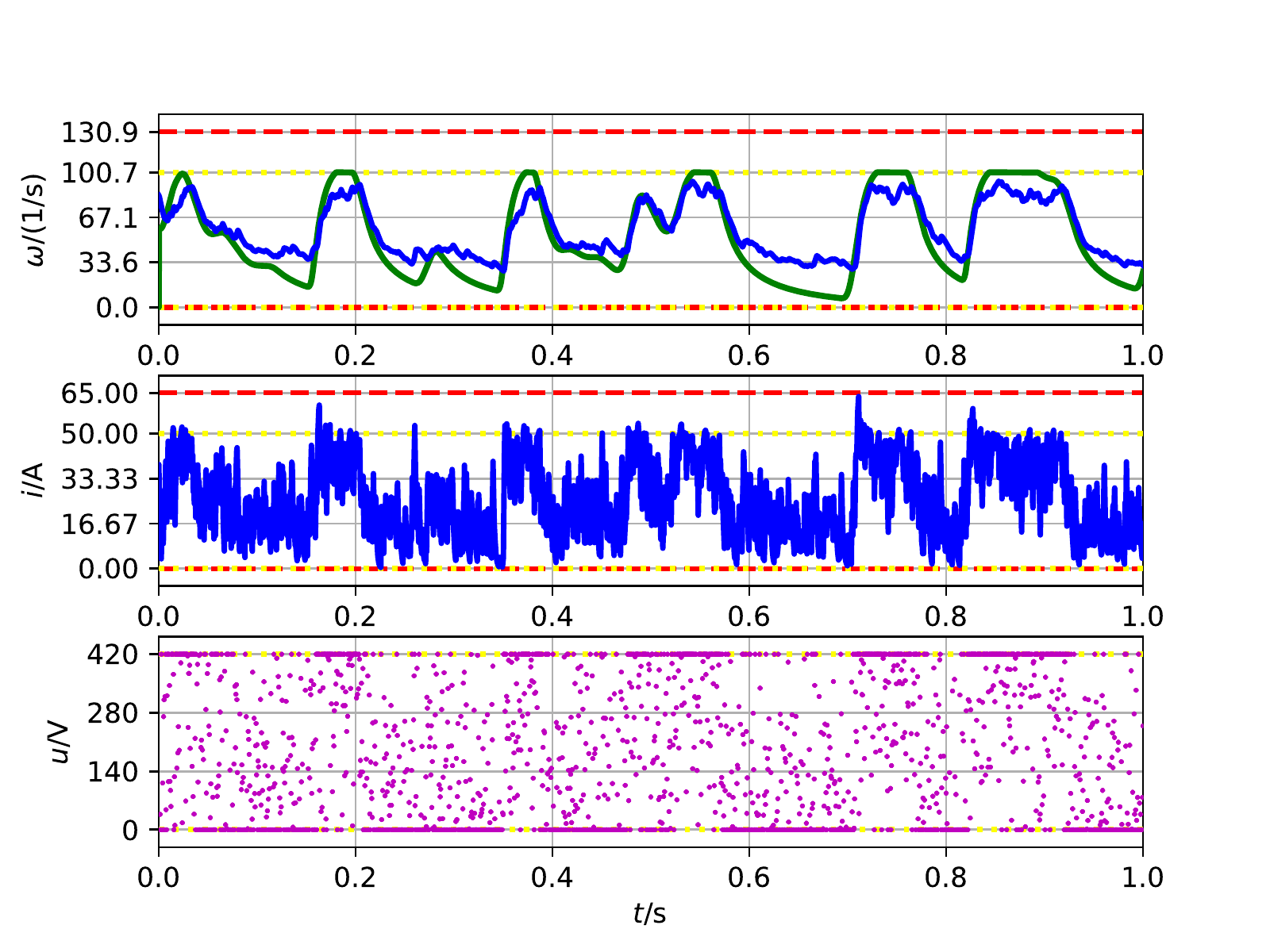}
	\caption{Trajectories of the \DDPG-agent (blue) and the input voltage (magenta) after $1\,000\,000$ training steps with a \MAE~of $0.0965$ are shown. The reference is depicted in green and the nominal values (dotted-yellow) and limits (dashed-red) are drawn.}
	\label{fig:LearningCurveRL2}
\end{figure}

\begin{figure}[!t]	
	\includegraphics[width=\columnwidth, trim=0cm 0.3cm 1cm 1.4cm, clip=true]{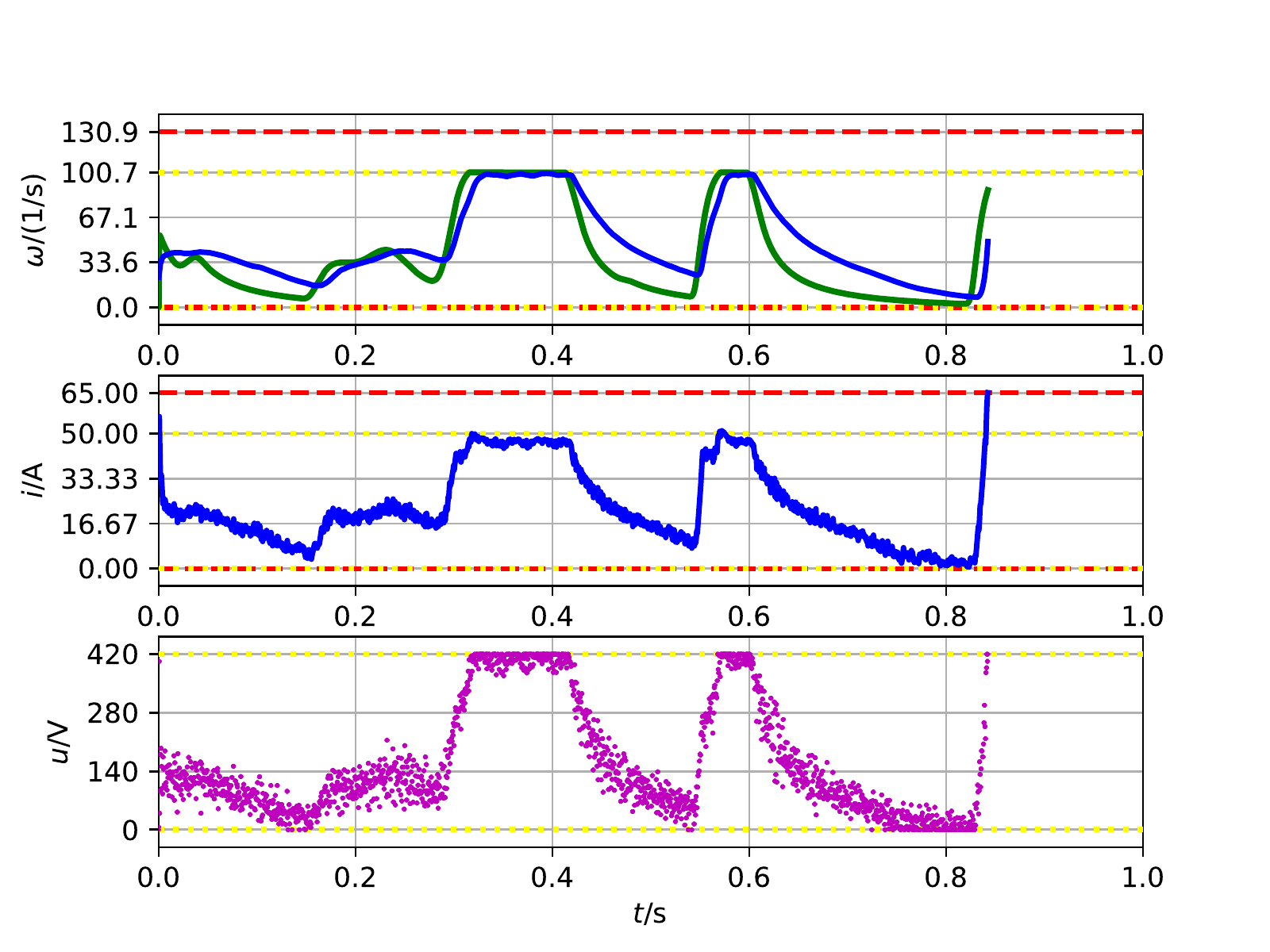}
	\caption{Trajectories after $5\,900\,000$ training steps with a \MAE~of $0.1122$ are plotted. The episode is stopped due to over-current. (colors cf. \figurename~\ref{fig:LearningCurveRL2})}
	\label{fig:LearningCurveRL3}
\end{figure}

\begin{figure}[!t]
	\includegraphics[width=\columnwidth, trim=0cm 0.9cm 1cm 2.1cm, clip=true]{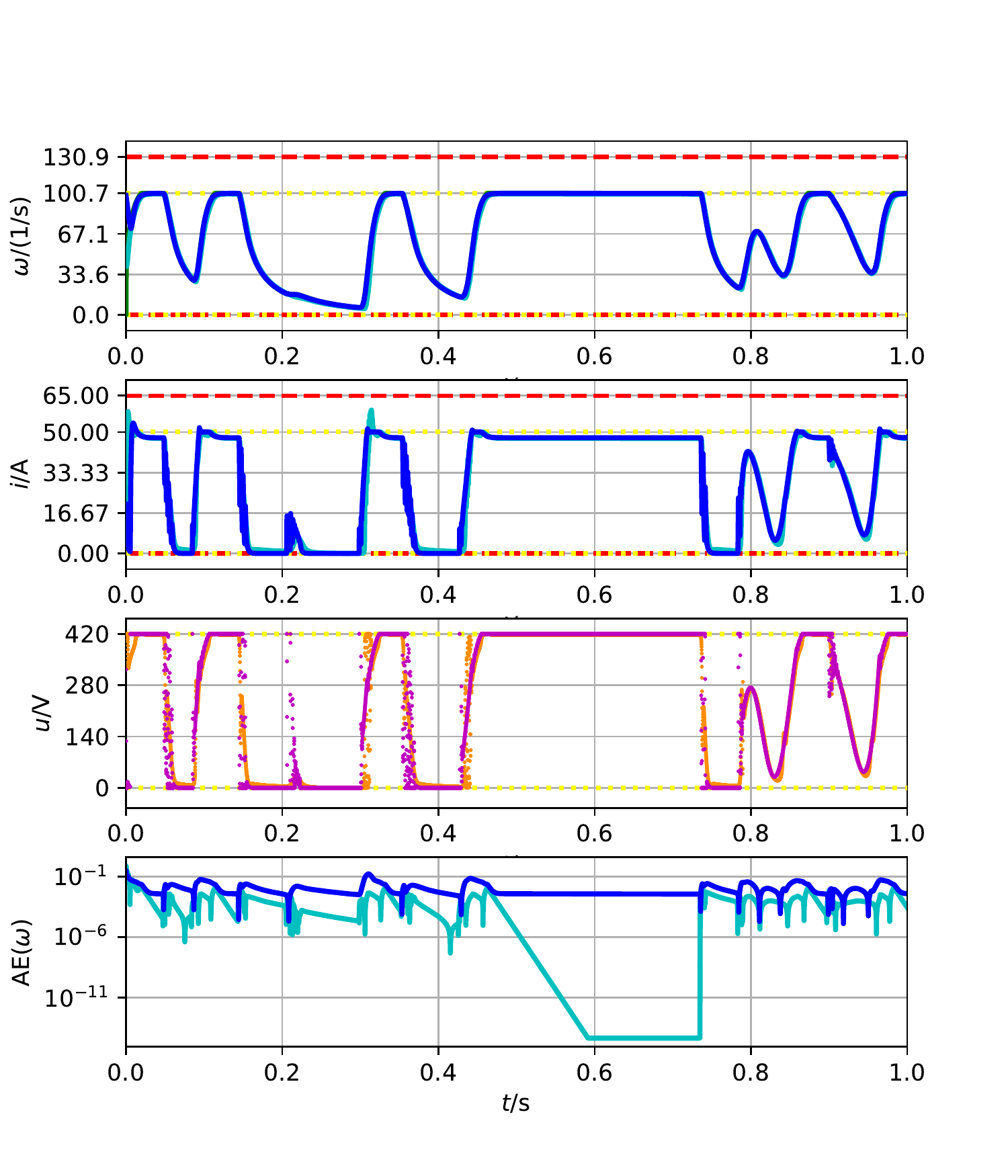}
	\caption[width=\textwidth]{Trajectories of learned ($7\,500\,000$ training steps) \RL-agent and the input voltage in comparison to a cascaded PI-controller (cyan) and its input voltage (orange) are drawn (colors cf. \figurename~\ref{fig:LearningCurveRL2}). In the bottom plot of, the absolute error of the \RL-agent and cascaded controller are plotted.}
	\label{fig:trajectories}
	\label{fig:ExampleTrajectoryError}
	\label{fig:ExampleTrajectory}
	
\end{figure}

The control behavior during the training and afterwards is visualized in \figurename~\ref{fig:LearningCurveRL2},~\ref{fig:LearningCurveRL3} and~\ref{fig:ExampleTrajectory}.
\figurename~\ref{fig:LearningCurveRL2} shows a control episode after about $1\,000\,000$ simulation steps.
The agent does not perform well and the MAE is $0.0965$.
The actions are very noisy, which can be seen in the input voltage plot. 
The Gaussian noise affects the actions at this point in the training process a lot.

A trajectory after about $5\,900\,000$ steps is plotted in \figurename~\ref{fig:LearningCurveRL3}.
The input voltage contains less noise, however, the reference tracking is worse than before, which is expressed by the \MAE~of $0.1122$.
Furthermore, this trajectory is prematurely stopped due to a limit violation.
The current limit is exceeded after the reference of the angular velocity is sharply increasing.
\RL-agents must learn to hold those limits to be applicable in real applications.

Trajectories of a learned agent after $7\,500\,000$ simulation steps are plotted in \figurename~\ref{fig:ExampleTrajectory}.
The angular velocity, the input voltage and the current are highlighted, similar to the dashboard in the toolbox.
Furthermore, the trajectories of a cascaded PI-controller for the same reference are included.
The \MAE~of the \DDPG-agent is $0.0133$, which is much smaller than in the two trajectories before, and the \MAE~of the PI-controller is even smaller with $0.0024$.
The dispersion over time of the absolute error between the angular velocity and its reference of the episode shown in \figurename~\ref{fig:ExampleTrajectory} can be seen in the bottom plot.
As expected, the error over time describes sudden jumps that align with jumps in the reference trajectory due to the low-pass behaviour of the system.
Moreover, it can be taken from the figure between \SI{0.4}{\second} and \SI{0.8}{\second}, that there is a small steady state error with the \DDPG-agent. 

The learned agent's average \MAE~over $100$ trajectories, given in Tab.~\ref{tab:MAEresults}, is in the same magnitude as the error of the cascaded controller.
This shows that the RL control approach for electric motors reaches control quality similar to a state-of-the-art controller, and that \RL~is a highly promising approach for electric motor control.
The control quality of the \DDPG-agent might be improved with an optimization of the \DDPG-parameters and architecture in future research.
The GEM toolbox supports this research with fast and easy creation of training environments for the RL-agents.



\begin{table}[h]
	\centering
	\renewcommand{\arraystretch}{1.3}
	\caption{\MAE~per step}
	\label{tab:MAEresults}
	\begin{tabular}{ccc}
		\toprule
		&DDPG&PI\\
		\midrule
		min of $100$ trajectories&$0.0009$&$0.0001$ \\
		mean of $100$ trajectories&$0.0631$&$0.0323$\\
		max of $100$ trajectories&$0.7037$&$0.6381$\\
		\bottomrule
	\end{tabular}
\end{table}

\section{Conclusion}
\label{sec:conclusion}
The novel open-source toolbox \gymem~for simulating electric motors for \RL-agents was presented. Details of the toolbox, as the combination of converter, motor and load as well as the rewards and the reference generation have been described. In an example, possible use cases of the toolbox are demonstrated and it also shows that \RL-agents and cascaded PI-controller are competitive. 
Several future research topics are of interest. General investigations about the competitiveness of \RL-agents and other control schemes are necessary. The hyperparameters of the \RL-agent can be optimized and the toolbox can be extended with an induction machine and more detailed converter models. Furthermore, the application on a real motor test bench will be part of future research to make motor control with \RL-agents useful for a wide range of applications. Fellow researchers are invited to work with the toolbox to develop their own \RL~electric motor control agents and to contribute to \gymem~in terms of model extensions or critical feedback.


\appendices

\begin{table}[h]
	\renewcommand{\arraystretch}{1.3}
	\centering
	\caption{List of variables used in the motor environments}
	\label{tab:variables}
	\begin{tabular}{*2cl} \hline
		\textbf{variable}&\textbf{meaning}\\
		\toprule
		\multicolumn{2}{c}{\textbf{general motor variables}}\\
		\midrule
		$\tau$&sampling time\\ \hline
		$\omega$, $\omega_{me}$ & electrical / mechanical angular velocity\\\hline
		$\varepsilon$, $\varepsilon_{me}$&electrical / mechanical rotor angle\\\hline
		$T$&torque from motor\\\hline
		$T_L(\omega)$&load torque\\\hline
		$J_{rotor}$, $J_{load}$&moment of inertia of motor / load\\\hline
		$\bm{u}_{in}$&input voltage\\\hline
		$u_{sup}$&supply voltage\\\hline
		$\bm{i}_{in}$&input current\\
		\toprule
		\multicolumn{2}{c}{\textbf{DC motor variables}}\\
		\midrule
		$R_A$, $R_E$&armature / excitation resistance\\\hline
		$L_A$, $L_E$&armature / excitation inductance\\\hline
		$L^\prime_E$&effective excitation inductance\\\hline
		$\mathit{\Psi}^\prime_E$&effective excitation flux\\\hline
		$i_A$,$i_E$&armature / excitation current\\\hline
		$u_A$,$u_E$&armature / excitation voltage\\\hline
		\toprule
		\multicolumn{2}{c}{\textbf{PMSM variables}}\\
		\midrule	
		$u_a$, $u_b$, $u_c$&phase voltage\\\hline
		$i_a$, $i_b$, $i_c$&phase currents \\\hline
		$i_{sd}$, $i_{sq}$&direct / quadrature axis current\\\hline
		$u_{sd}$, $u_{sq}$&direct / quadrature axis voltage\\\hline
		$R_s$&stator resistance\\\hline
		$L_d$, $L_q$&direct / quadrature axis inductance\\\hline
		$p$&pole pair number\\\hline
		$\mathit{\Psi}_p$& permanent linked rotor flux\\	
		\bottomrule
	\end{tabular}
\end{table}

\ifCLASSOPTIONcaptionsoff
  \newpage
\fi

\bibliographystyle{IEEEtran}
\bibliography{PaperRLTraueBook}

\begin{thebibliography}{10}
\providecommand{\url}[1]{#1}
\csname url@samestyle\endcsname
\providecommand{\newblock}{\relax}
\providecommand{\bibinfo}[2]{#2}
\providecommand{\BIBentrySTDinterwordspacing}{\spaceskip=0pt\relax}
\providecommand{\BIBentryALTinterwordstretchfactor}{4}
\providecommand{\BIBentryALTinterwordspacing}{\spaceskip=\fontdimen2\font plus
\BIBentryALTinterwordstretchfactor\fontdimen3\font minus
  \fontdimen4\font\relax}
\providecommand{\BIBforeignlanguage}[2]{{%
\expandafter\ifx\csname l@#1\endcsname\relax
\typeout{** WARNING: IEEEtran.bst: No hyphenation pattern has been}%
\typeout{** loaded for the language `#1'. Using the pattern for}%
\typeout{** the default language instead.}%
\else
\language=\csname l@#1\endcsname
\fi
#2}}
\providecommand{\BIBdecl}{\relax}
\BIBdecl

\bibitem{Linder.2010}
A.~Linder, R.~Kanchan, P.~Stolze, and R.~Kennel, \emph{Model-Based Predictive
  Control of Electric Drives}, 1st~ed.\hskip 1em plus 0.5em minus 0.4em\relax
  G{\"o}ttingen: {Cuvillier Verlag}, 2010.

\bibitem{Gorges.2017}
D.~G{\"o}rges, ``Relations between model predictive control and reinforcement
  learning,'' \emph{IFAC-PapersOnLine}, vol.~50, no.~1, pp. 4920--4928, 2017.

\bibitem{Krizhevsky.2017}
A.~Krizhevsky, I.~Sutskever, and G.~E. Hinton, ``Imagenet classification with
  deep convolutional neural networks,'' \emph{Communications of the ACM},
  vol.~60, no.~6, pp. 84--90, 2017.

\bibitem{Mnih.2015}
{V. Mnih et al.}, ``Human-level control through deep reinforcement learning,''
  \emph{Nature}, vol. 518, no. 7540, pp. 529--533, 2015.

\bibitem{Lillicrap.992015}
\BIBentryALTinterwordspacing
{T. Lillicrap et al.}, ``Continuous control with deep reinforcement learning,''
  2015. [Online]. Available: \url{arXiv:1509.02971}
\BIBentrySTDinterwordspacing

\bibitem{Silver.2017}
{D. Silver et al.}, ``Mastering the game of go without human knowledge,''
  \emph{Nature}, vol. 550(7676), pp. 354--359, 2017.

\bibitem{Panyakaew.2018}
S.-N. Panyakaew, P.~Inkeaw, J.~Bootkrajang, and J.~Chaijaruwanich, ``Least
  square reinforcement learning for solving inverted pendulum problem,'' in
  \emph{2018 3rd International Conference on Computer and Communication
  Systems}.\hskip 1em plus 0.5em minus 0.4em\relax Piscataway, NJ: {IEEE
  Press}, 2018, pp. 16--20.

\bibitem{Hesse.2018}
M.~Hesse, J.~Timmermann, E.~H{\"u}llermeier, and A.~Tr{\"a}chtler, ``A
  reinforcement learning strategy for the swing-up of the double pendulum on a
  cart,'' \emph{Procedia Manufacturing}, vol.~24, pp. 15--20, 2018.

\bibitem{Fremaux.2013}
N.~Fr{\'e}maux, H.~Sprekeler, and W.~Gerstner, ``Reinforcement learning using a
  continuous time actor-critic framework with spiking neurons,'' \emph{PLoS
  computational biology}, vol.~9, no.~4, p. e1003024, 2013.

\bibitem{Glavic.2017}
M.~Glavic, R.~Fonteneau, and D.~Ernst, ``Reinforcement learning for electric
  power system decision and control: Past considerations and perspectives,''
  \emph{IFAC-PapersOnLine}, vol.~50, no.~1, pp. 6918--6927, 2017.

\bibitem{Schenke.}
M.~Schenke, W.~Kirchg{\"a}ssner, and O.~Wallscheid, ``Controller design for
  electrical drives by deep reinforcement learning: a proof of concept,''
  \emph{IEEE Transactions on Industrial Informatics (submitted)}, 2019.

\bibitem{Mnih.19.12.2013}
V.~Mnih, K.~Kavukcuoglu, D.~Silver, A.~Graves, I.~Antonoglou, D.~Wierstra, and
  M.~Riedmiller, ``Playing atari with deep reinforcement learning,''
  \emph{NIPS}, 2013.

\bibitem{Brockman.652016}
\BIBentryALTinterwordspacing
G.~Brockman, V.~Cheung, L.~Pettersson, J.~Schneider, J.~Schulman, J.~Tang, and
  W.~Zaremba, ``Open{AI} {G}ym,'' 2016. [Online]. Available:
  \url{arXiv:1606.01540}
\BIBentrySTDinterwordspacing

\bibitem{Plappert.2016}
\BIBentryALTinterwordspacing
M.~Plappert, ``keras-rl,'' 2016. [Online]. Available:
  \url{https://github.com/keras-rl/keras-rl}
\BIBentrySTDinterwordspacing

\bibitem{tensorforce}
\BIBentryALTinterwordspacing
A.~Kuhnle, M.~Schaarschmidt, and K.~Fricke, ``Tensorforce: a tensorflow library
  for applied reinforcement learning,'' 2017. [Online]. Available:
  \url{https://github.com/tensorforce/tensorforce}
\BIBentrySTDinterwordspacing

\bibitem{baselines}
\BIBentryALTinterwordspacing
P.~Dhariwal, C.~Hesse, O.~Klimov, A.~Nichol, M.~Plappert, A.~Radford,
  J.~Schulman, S.~Sidor, Y.~Wu, and P.~Zhokhov, ``Open{{AI}} baselines,'' 2017.
  [Online]. Available: \url{https://github.com/openai/baselines}
\BIBentrySTDinterwordspacing

\bibitem{Sutton.2018}
R.~S. Sutton and A.~Barto, \emph{Reinforcement learning: An introduction},
  second edition~ed., ser. Adaptive computation and machine learning.\hskip 1em
  plus 0.5em minus 0.4em\relax Cambridge, MA and London: {The MIT Press}, 2018.

\bibitem{Bocker.08.01.2018}
J.~B{\"o}cker, \emph{Electrical Drive Systems (in~{{G}}erman)}, Paderborn
  University, 2018.

\bibitem{Bocker.11.07.2018}
------, \emph{Controlled Three-Phase Drives}, Paderborn University, 2018.

\bibitem{Chiasson.2005}
J.~Chiasson, \emph{Modeling and High-Performance Control of Electric
  Machines}.\hskip 1em plus 0.5em minus 0.4em\relax Hoboken, NJ, USA: {John
  Wiley {\&} Sons, Inc}, 2005.

\bibitem{Prof.Dr.JoachimBocker.Sommersemester2019}
J.~B{\"o}cker, \emph{Power Electronics}, Paderborn University, 2019.

\bibitem{Butcher.2008}
J.~C. Butcher, \emph{Numerical methods for ordinary differential
  equations}.\hskip 1em plus 0.5em minus 0.4em\relax Hoboken, N.J: Wiley, 2008.

\bibitem{.2009}
D.~Schr{\"o}der, \emph{Elektrische Antriebe - Regelung von
  Antriebssystemen}.\hskip 1em plus 0.5em minus 0.4em\relax Berlin, Heidelberg:
  {Springer Berlin Heidelberg}, 2009.

\end{thebibliography}
\end{document}